\documentclass[journal]{IEEEtran}
\ifCLASSINFOpdf

\else
\fi
\usepackage{graphicx,amsmath,amssymb,cite}
\usepackage{subfigure}
\usepackage{color}
\usepackage{amsfonts}
\usepackage{amsmath}
\usepackage{graphicx}
\usepackage{amssymb}
\usepackage{stfloats}
\usepackage{mathrsfs}
\usepackage{algorithm}
\usepackage{algorithmic}

\begin{document}

\title{Spatial Modulation: an Attractive Secure Solution to Future Wireless Network}

\author{Feng Shu, Lin Liu, LiLi Yang, Xinyi Jiang, Guiyang Xia, Yuanyuan Wu,\\Xianpeng Wang, Shi Jin, Jiangzhou Wang, and Xiaohu You.
\thanks{This work was supported in part by the National Natural Science Foundation of China (Nos. 62071234 and 61771244)(Corresponding authors: Feng Shu and Xianpeng Wang).}
\thanks{Feng Shu, Guiyang Xia, Yuanyuan Wu, and Xianpeng Wang are with the School of Information and Communication Engineering, Hainan University, Haikou, 570228, China (e-mail: shufeng0101@163.com; xiaguiyang@njust.edu.cn; wyuanyuan82@163.com; wxpeng2016@hainanu.edu.cn).}
\thanks{Lin Liu, LiLi Yang, Xinyi Jiang, and Feng Shu are with School of Electronic and Optical Engineering, Nanjing University of Science and Technology, Nanjing, 210094, China (e-mail: liulin\_sp60@njust.edu.cn; yanglily@njust.edu.cn; jiangxinyi0313@163.com).}
\thanks{Shi Jin and Xiaohu You are with the National Mobile Communication Research Lab, Southeast University, Nanjing 210096, China (e-mail: jinshi@seu.edu.cn; xhyu@seu.edu.cn).}
\thanks{Jiangzhou Wang is with the School of Engineering and Digital Arts, University of Kent, Canterbury CT2 7NT, U.K (e-mail: j.z.wang@kent.ac.uk).}
}
\maketitle
\begin{abstract}
  As a green and secure wireless transmission method, secure spatial modulation (SM) is becoming a hot research area. Its basic idea is to exploit both the index of activated transmit antenna and amplitude phase modulation signal to carry messages, improve security, and save energy. In this paper, we review its crucial challenges: transmit antenna selection (TAS), artificial noise (AN) projection, power allocation (PA) and joint detection at the desired receiver. As the size of signal constellation tends to medium-scale or large-scale, the complexity of traditional maximum likelihood detector becomes prohibitive. To reduce this complexity, a low-complexity maximum likelihood (ML) detector is proposed. To further enhance the secrecy rate (SR) performance, a deep-neural-network (DNN) PA strategy is proposed. Simulation results show that the proposed low-complexity ML detector, with a lower-complexity, has the same bit error rate performance as the joint ML method while the proposed DNN method strikes a good balance between complexity and SR performance.
\end{abstract}

\section{Secure spatial modulation and deep learning}
Spatial modulation (SM) concept was first proposed by Chau and Yu in \cite{Space_modulation2001}. Its main idea is to carry additive bit information via antenna indices. In \cite{Mesleh2008Spatial}, the authors made a systematic and in-depth investigation of SM. Until now, the basic principle of SM was also extended to index modulation. SM exploits both the index of activated transmit antenna and amplitude phase modulation signal to carry messages. Different from Bell Laboratories Layer Space-Time (BLAST) and space time coding (STC), SM may strike a good balance between spatial multiplexing and diversity and is called the third way between BLAST and STC. Compared to BLAST and STC, SM has a merit of high energy efficiency due to the use of less active radio frequency (RF) chains. Thus, it is also a green wireless transmission technique.

Wireless communication is usually prone to passive eavesdropping and active malicious attack due to its open broadcast characteristics. Although there is a series of mature encryption algorithms in the upper layer of network protocol, it is still possible to be broken in wireless communication if the eavesdropper has a strong computational ability. To address this problem, the physical layer security (PLS) technology becomes an inevitable choice to work with traditional encryption methods to provide a double layer protection on confidential message (CM), and enhances wireless security from the perspective of information theory. PLS has been extensively studied in \cite{PLSchallenges2015}. PLS actually provides an incremental  guarantee for the future personal privacy protection and information network security.

Recently, secure modulation has emerged as an special form of multiple-input-multiple-output (MIMO). It  is mainly composed of two categories: directional modulation (DM) and secure SM (SSM). DM, with the help of artificial noise (AN), can securely deliver CM to desired user in line-of-sight channel by beamforming, and  is unsuitable for fading channels. Conversely, SM is naturally suitable for fading channel.

Transmitting CM via SM is an attractive and very important issue \cite{Feng2018Two,xia2019antenna}. In \cite{Feng2018Two}, the authors have made a wide and in-depth investigation of transmit antenna selection (TAS) methods in SSM systems. Then, two high-performance TAS schemes: leakage-based and  maximum secrecy rate (SR), have been proposed to improve the SR performance, and the generalized Euclidean distance-optimized antenna selection (EDAS)  method has been extended to provide a secure transmission.  In \cite{xia2019antenna}, an active antenna-group selection was proposed to maximize the average SR for limited active antenna pattern and finite-alphabet inputs.

In SSM,  how to construct a proper AN projection matrix has an important impact on the SR performance. In \cite{Wu2015Secret, xia2019antenna}, AN was projected  onto the null-space of the desired channel to improve the SR performance. Here, the major benefit of this scheme is the fact that the AN projection matrix has a closed-form expression and is of low-complexity. However, such a scheme might result in some secrecy performance loss.
\begin{figure*}[htbp]
\centering
\includegraphics[width=0.97\textwidth]{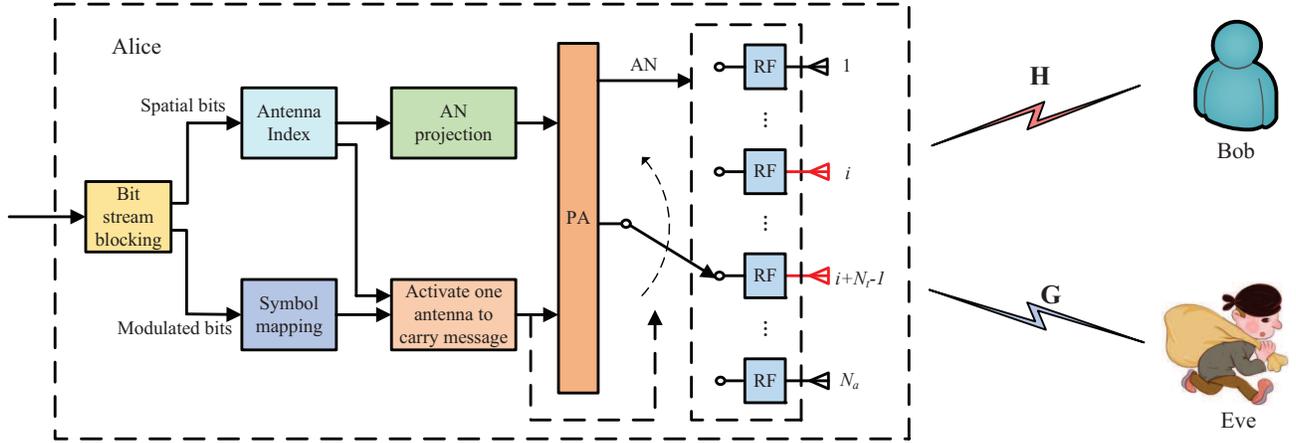}
\centering
\caption{Schematic diagram of SSM network.}\label{Spatial modulation_fig1}
\end{figure*}

Intelligent communication has been considered as one of the mainstream directions of the coming future development of wireless communication. Its basic idea is to introduce intelligent elements into all layers of wireless networks, so as to realize the organic integration of wireless networks and artificial intelligence technology, and greatly improve the efficiency and performance of wireless networks. The existing research results have concentrated on the application layer, physical layer, and the network layer. The main idea is to introduce machine learning, especially deep learning (DL), into wireless resource management, channel decoding, and other fields.

The DL has been successfully and widely applied in many fields such as computer vision, natural language processing, speech recognition, etc., and has achieved great success. Due to the new features and challenges of future wireless communication, such as complex scenes with unknown channel models, high-speed and accurate processing requirements, many scholars have introduced DL into the physical layer of wireless communication \cite{C.Jiang2017Machine}. In the physical layer, there is a new tendency of combining wireless transmission and DL. In \cite{He2018Deep}, the authors considered channel estimation for millimeter-wave massive MIMO systems. An approximate messaging network based on learning denoising was proposed for channel estimation, which can learn channel structure and estimate channel from a large amount of training data. In \cite{Nachmani2016Learning} a new framework was proposed for integrating large-scale MIMO and DL to address the problem of channel estimation and direction of arrival estimation.

In Fig.~\ref{Spatial modulation_fig1}, a conventional SSM system  is presented. In this figure, four important tools including TAS, beamforming of CMs, AN projection, and power allocation (PA) are fully employed to achieve a high-performance SSM. In such a system, at desired transmitter, the PA of maximizing SR is a hard problem considering the fact that the expression of SR has no closed-form. In other words, SR is a non-linear function of PA factor. Exhaustive search (ES) can achieve the optimal SR performance with a sufficient small bin width. To reduce the computational complexity and approach the optimal SR performance, a deep neural network (DNN) based PA strategy is proposed to implement PA between AN and CM given the known beamforming vector and AN projection matrix. With a slight SR performance loss, the proposed DNN-based PA method is of lower-complexity than ES.

\section{System model and transmit antenna selection}
Consider a typical SSM system as shown in Fig.~\ref{Spatial modulation_fig1}, where the transmitter (Alice) is equipped with $N_a$ transmit antennas (TAs). According to the nature of SM, when the number of TAs is not a power of two, $N_t = 2^{\left\lfloor\log _{2}^{N_a}\right\rfloor}$ out of $ N_a $ TAs have to be selected for mapping binary bits to the antenna index. Moreover, $\log_{2}^{M} $ bits are used to form a constellation symbol and $M$ is the size of the adopted signal constellation. As a result, the achievable spectral efficiency arrives at $\log _{2} N_{t}+\log _{2} M $ bits per channel use.

Appropriately selecting out an active antenna group is capable of improving the security performance of SM systems. As a matter of fact, there exist a number of TAS methods for enhancing the secrecy performance of SSM, such as random, leakage \cite{Feng2018Two}, and generalized EDAS. For the leakage-based TAS strategies,  the signal-to-leakage-and-noise ratio (SLNR) of CM from each transmit antenna is calculated and sorted, where the SLNR is defined by the ratio of the receive signal power at Bob to the sum of the receive power of CM at Eve, receive AN power and channel noise variance. Then a low-complexity sorting algorithm places the values of all the SLNRs in a decreasing order. Upon choosing the antennas associating with the top $N$ SLNRs, the so-called Max-SLNR method is established \cite{Feng2018Two}. The Max-SLNR has a ability of approaching the near-optimal SR performance with a low computational complexity.

From the perspective of the decoding performance at a receiver, generalized EDAS performs best in terms of bit error rate (BER). The generalized EDAS method aims for selecting out a TAS pattern by maximizing the minimum Euclidean distance over the desired channel or minimizing the minimum Euclidean distance over eavesdropping channel, the core principle is that the minimum distance has a direct impact of BER performance.
\begin{figure}[htbp]
\centering
\includegraphics[width=0.49\textwidth]{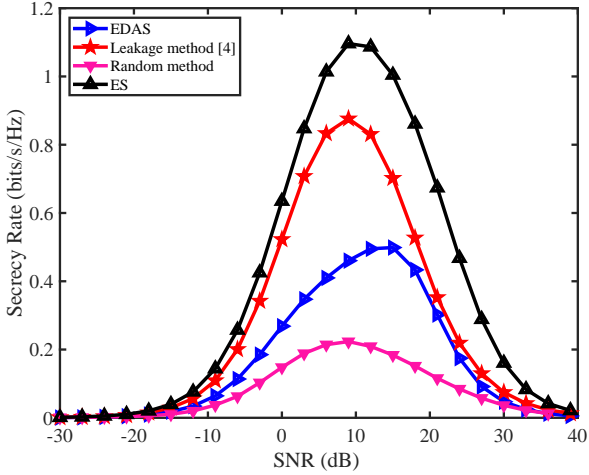}
\caption{Comparison of SR performance of various TAS methods.}\label{Comparison-of-SR-No-AN}
\end{figure}

Fig.~\ref{Comparison-of-SR-No-AN} shows a comparison of the SR performances of the optimal ES, Max-SLNR, EDAS, and random method with $N_a\!=\!15$, $N_t\!=\!8$, $N_b\!=\!N_e\!=\!2$. From this figure, it can be observed that  the four methods have a decreasing order in the SR performance as follows: ES, Max-SLNR, EDAS, and random method. Additionally, we also find an interesting result: all the SR curves first go up as hills, then reach their peaks, and finally go down hills as the SNR increases. In other words, all the SR curves have crest values, and can be regarded as concave functions of the SNR.

\section{Proposed DNN-based PA strategy}
How to allocate power among CM and AN will have a dramatic impact on the SR performance. PA, as an efficient way to enhance security in SM system, has been investigated in \cite{shu2019high}, where two high-performance PA strategies were proposed to achieve substantial SR gains over existing PA methods. The optimal PA factor between signal and interference transmission can be determined by exhaustive search for SM system. However, there is no closed-form SR expression in discrete-input continuous-output memoryless channels, which results in a high computational complexity to complete SR. In addition, small search step size also leads to a high computational complexity of ES. To reduce the computational complexity, an approximate SR (ASR) expression is used instead of exact SR. Then, gradient descent (GD) algorithm is adopted to solve the PA factor.

\begin{figure}[htbp]
\centering
\includegraphics[width=0.49\textwidth]{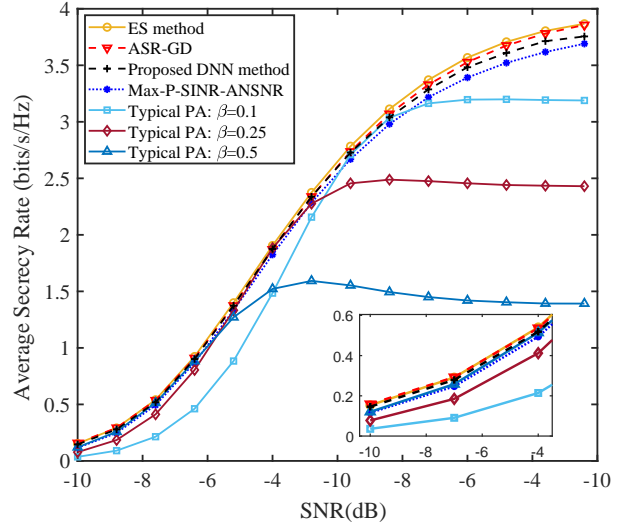}
\caption{Comparison of SR performance of various PA strategies.}\label{Comparison-of-PA}
\end{figure}

To make a complete comparison among existing methods and the proposed DNN in this section, simulation parameters are set as follows: $N_a\!=\!N_t\!=\!4$, $N_b\!=\!N_e\!=\!2$. Fig.~\ref{Comparison-of-PA} makes a comparison of several typical PA strategies: ES, fixed, ASR-GD, Max-P-SINR-ANSNR, and proposed DNN.
Comparing the three methods with fixed PA factors, it can be seen that the SR at $\beta=0.5$ is the lowest one, and $\beta=0.1$ is the highest one in the value of high SNR. This is because when the SNR is high, both Bob and Eve have a very good quality of channel, and a large portion of transmit power may be allocated to AN to disturb eavesdropper, so as to obtain a high security performance. Correspondingly, when the SNR is low, a high portion of transmit power should be allocated to CM to improve Bob's reachable rate, so as to achieve a high security performance.

As shown in Fig.~\ref{Comparison-of-PA}, the SR performance of ASR-GD is close to that of ES method. However, to evaluate the expression of ASR is still high-complexity. The authors in \cite {shu2019high} analyzed the problem in terms of the power of the received signal and noise, and proposed a novel PA strategy called Max-P-SINR-ANSNR, where 'P' is short for product, and 'ANSNR' stands for AN-to-signal-plus-noise ratio, presented a closed-form expression for the PA factor. As shown in Fig.~\ref{Comparison-of-PA}, the SR performance of the proposed Max-P-SINR-ANSNR is close to that ASR-GD, but with extremely low complexity.
\begin{figure}[htbp]
\centering
\includegraphics[width=0.49\textwidth]{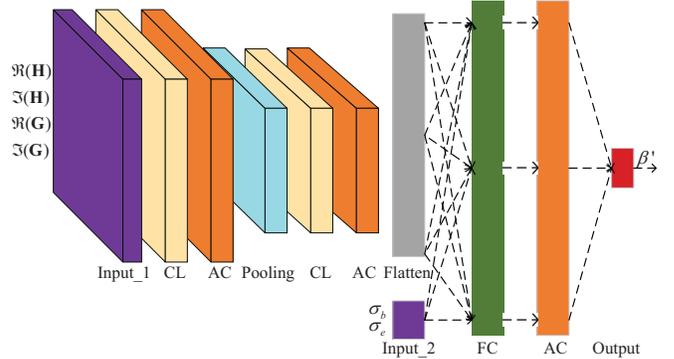}
\caption{Diagram of the deep neural network.}\label{DNN1}
\end{figure}

To reduce the computational complexity and approach the optimal SR performance, we proposed a DNN-based PA strategy. The key idea is to treat the input and output of  ES algorithm as an unknown nonlinear mapping and use a DNN to approximate it.
The network structure is illustrated in Fig.~\ref{DNN1}. More specifically, the network consists of two input layers, an output layer, convolutional (CL) layers, activation (AC) layers, pooling layer and fully-connected (FC) layer. The input\_1 is 3-dimensional matrix, which consists of the real and imaginary parts of the desired channel matrix and eavesdropping channel matrix.
Then the features are extracted through operations such as CL, AC, pooling, CL and AC.
Next, the extracted features are flattened and combined with the noise variance as the input of the FC layer.
Finally, the DNN outputs the predicted value of PA.
\begin{figure*}[htbp]
	\centering
	\includegraphics[width=1.0\textwidth, angle=0]{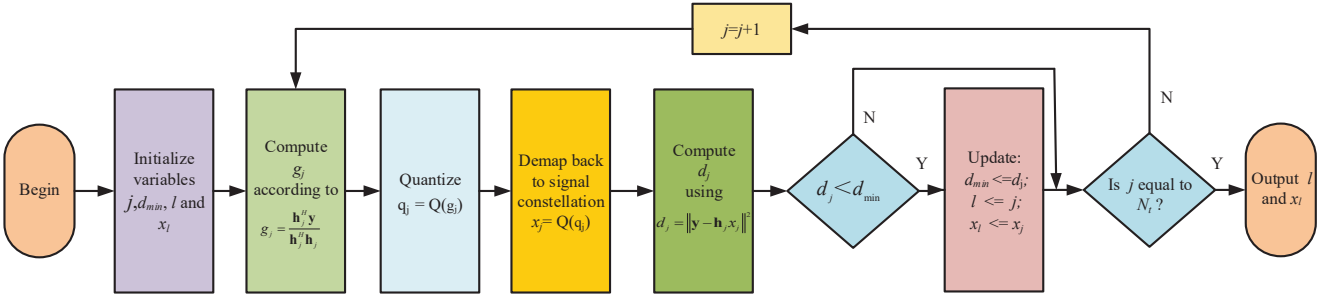}
	\caption{Block diagram of the proposed low-complexity ML detector.}\label{dnn_fig01}
\end{figure*}

In addition to the influence of network structure, the performance of DNN largely depends on the constructed training data set and its training method. First, the channel state information (CSI) of the desired and eavesdropping users obeying Rayleigh distribution are randomly generated, and then the optimal PA factor is solved by ES method as a label. The set of CSI and PA factor are used as the training data set, and then Adam optimizer is used to train DNN off-line. The trained DNN is used to predict the PA factor based on the newly input CSI about the desired user and the eavesdropping user. As shown in Fig.~\ref{Comparison-of-PA}, the SR performance of the proposed DNN is close to that of ASR-GD algorithm, and  strikes a good balance between complexity and performance.

\section{Proposed low-complexity ML detector}
In \cite{4601434}, the authors proposed an optimal joint ML detector. Here, Bob is assumed to have the perfect knowledge of \textbf{H} and \textbf{S} where \textbf{H} denotes the channel from Alice to Bob and \textbf{S} is the associated antenna selection matrix.  This joint ML detector can reach an optimal BER performance but it needs exhaustive search among all possible antennas and symbols to infer the most likely solution, which requires a  high computational complexity, especially in large-scale signal constellation. To reduce computational complexity,  a sub-optimal method with low-complexity in \cite{4149911} was proposed but had a far worse BER performance than joint ML. To address this dilemma, a new low-complexity ML detector is proposed by us to approach the ML performance in this paper. It utilizes the CSI and received signal to detect the symbol, and then combines the symbol and CSI to estimate the antenna index. The corresponding flowchart for the proposed method is plotted in Fig.~\ref{dnn_fig01}.

The whole process for our proposed method can be described as follows.  First, we  initialize the estimated symbol $x_l$, estimated antenna index $l$, and the minimum Euclidean distance $d_{min}$, where  $\textbf{h}_j$ is the $\emph{j}$th column of $\textbf{H}$ and $\textbf{y}$ represents the received signal vector. For all  $N_t$ transmit antennas, we traverse these transmit antennas to calculate the corresponding $g_j$, which is used to demodulate the transmit symbol $x_j$. $Q(\cdot)$ and $D(\cdot)$ represent the quantization and demapping functions, respectively. The latter maps the quantized value to the nearest constellation points.  Then  the Euclidean distance $d_j$  between receiver signal $\textbf{y}$ and the production of assumed channel $\textbf{h}_j$ and estimated symbol $x_m$ is computed. If $d_j$ is less than the predefined $d_{min}$, it means this process can terminate and we will update all the original values. If not,  we will move to the next antenna index until all $N_t$ transmit antennas are traversed.

To  make a comparison among the computation complexities of  the low-complexity ML detector proposed by us, ML and suboptimal methods, we take the number of complex multiplications (CMs) as a performance metrics. As in these two processes, the numbers of complex additions for all methods are identical, therefore, we omit the number of additions here. Using the results in \cite{4601434}, their complexities are follows: $\emph{C}_{ML}\!=\!2N_tN_r\!+\!2N_tM\!+\!M$ CMs, $\emph{C}_{Proposed}\!=\!2N_tN_r\!+\!N_t\log_2M\!+\!2N_t$ CMs, and $\emph{C}_{Suboptimal}\!=\!2N_tN_r\!+\!N_t\!+\!M$ CMs. Obviously, the proposed method has much lower computational complexity than the ML method as the number of constellation points tends to large-scale.
\begin{figure}[htbp]
\centering
\includegraphics[width=0.5\textwidth]{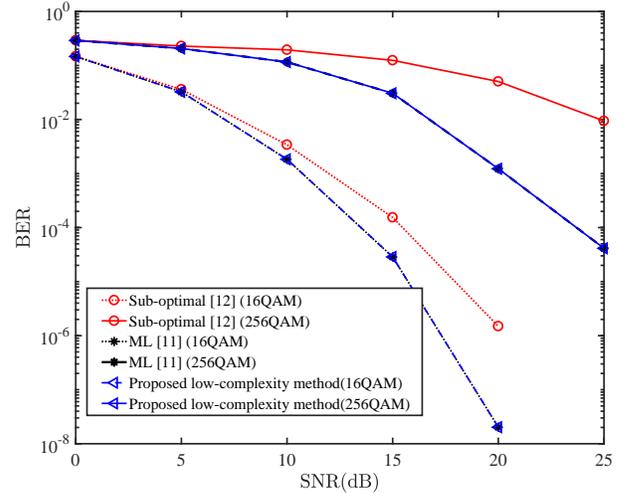}
\caption{Comparison of BER performance in SM system employing different detection.}\label{frame}
\end{figure}

To evaluate the BER performance of the proposed method, simulation parameters are chosen as follows: the transmit power being 4W, $N_t$=4, $N_r$=4, modulation 16 QAM or 256 QAM modulation. Fig.~\ref{frame} plots the curves of BER versus SNR for three detectors. From both 16 QAM and 256 QAM,  it is clearly seen that the proposed low-complexity ML detector achieves the same BER performance as the joint ML and far better than suboptimal method. Specially, the proposed detector harvests a SNR gain of 2.5dB at BER=$10^{-6}$ over the sub-optimal detection for 16 QAM. As the SNR increases, the BER performance gain increases gradually. For 256 QAM, the SNR gain becomes significant. At BER=$10^{-2}$, the proposed detector achieves a SNR gain of 8dB over sub-optimal method.

\section{Spatial modulation and intelligent reflecting surface}
Communications with intelligent reconfigurable surface (IRS) have been regarded as a promising candidate technology for future wireless networks. An IRS is an electromagnetic two-dimensional surface, composed of a large number of low-cost nearly-passive reconfigurable reflecting elements. Equipped with a smart controller, the IRS is able to intelligently adjust the phases of incident electromagnetic waves to increase the received signal energy, expand the coverage region, and alleviate interference, so as to enhance the communication quality of the wireless networks.

As mentioned above, SM is a special MIMO technology which activates one transmit antenna with one transmit antenna and exploits the index of the active antenna for information transfer. The undeniable potential of both SM and IRS based communication schemes has been the main motivation of this part.

 The concept of IRS-assisted communications was first brought to the realm of SM in \cite{Basar}. In \cite{Yanwenjing}, the authors applied SM principle to the IRS by adjusting the ON (active) and OFF (inactive) status of each reflecting element. Therefore, the IRS can deliver additional information by adopting SM on the index of the reflecting elements. In \cite{Basar}, the authors investigated the IRS-aided receive SM (RSM) technique. Inspired by \cite{Basar}, the authors in \cite{Mateng} extended its structure to combine the transmit and receive antenna indices for joint spatial modulation by shaping the reflecting beam with IRS. It is worth mentioning that conventional SM cannot combine transmit SM and RSM at the same time, due to the limitation of a single activated transmit antenna.

 In \cite{Basar,Yanwenjing,Mateng}, it has been shown that the IRS-assisted SM system outperforms, in terms of achievable rate and BER, the conventional SM system. Additionally, there is scarce existing literature studying the SR performance of IRS-assisted SSM system, which might be an potential way to make a significant improvement in SR performance.

\section{Open problems}
 There are still so many open existing problems  to be addressed in SSM. In what follows,  several important ones of them are summarized:
 \begin{enumerate}
  \item If Eve works at a full-duplex model, she will become  an active eavesdropper, i.e., a fixture of Mallory plus Eve. In such a situation, it is particularly important  to optimize the design of the transmitter at Alice in order to reduce the effect of jamming from Eve and at the same time achieve a feasible performance. This is a hard task. For Bob, the receive beamforming scheme  should be well designed to combat the jamming.

  \item In the presence of CSI measurement errors, how to construct robust beamforming, PA, and TAS by taking the statistical property of CSI error into account requires a great effort. In particular, the first task is to derive the SR closed-form expression or approximate expression in such a situation. This will pave a successful way for robust beamforming, PA, and TAS.

   \item  As the number of  transmit antennas at  transmitter tends to massive, the circuit cost and complexity becomes prohibitive for future practical applications of SM. Introducing hybrid analog and digital MIMO structure into SM is an efficient way to reduce the detection complexity and exploit the spatial diversity gain from MIMO. This will make a good balance among circuit cost, computational complexity and secure performance.

  \item Currently, IRS is popular, combining IRS with SM is a new trend. This will open a new life for  SM. How to establish the system model of secure IRS-aided SM is changeling. Based on this, it is also nontrivial to derive the closed-form expression and upper bound of SR. Finally, it is extremely important for how to optimize the choice of IRS elements under the criterion of maximizing SR.

\end{enumerate}

\section{Conclusion}
In this paper, the great potential of SSM has been highlighted as a key secure tool for future wireless networks such as vehicular communications, internet of things, unmanned aerial vehicle, smart transportation, and satellite communications. We reviewed its key techniques: beamforming, TAS, PA, and detection at desired receiver. A DNN-based PA was proposed to achieve the SR performance close to the optimal one. Then, a low-complexity ML  detector is proposed to achieve the optimal BER performance of joint ML. Also, several new open important future research problems are raised for SSM. Finally, it is pointed out that SSM will have  several diverse promising applications in the coming future.

\ifCLASSOPTIONcaptionsoff
  \newpage
\fi
\bibliography{IEEEfull,reference}
\bibliographystyle{IEEEtran}
\begin{IEEEbiographynophoto}\\
FENG SHU is a professor with the School of Information and Communication Engineer at Hainan University, Haikou, China.  In 2020, he was awarded with the Leading talents of Hainan Province. Also, he has  been awarded with Mingjian Scholar Chair Professor and Fujian Hundred-talents Program in Fujian Province, China.  He has published more 200 journal  papers  and with more than 150 SCI-indexed papers and more than 100 IEEE journal papers. Now, he serve as an editor for IEEE journals IEEE Systems Journal and IEEE Wireless Communications Letters. His research interests include  massive MIMO (especially directional and spatial modulations),  DOA measurements, wireless location, and machine learning for mobile communications.
\end{IEEEbiographynophoto}
\begin{IEEEbiographynophoto}\\
LIN LIU is a postgraduate student in the School of Electronic and Optical Engineering at Nanjing University of Science and Technology, Nanjing, China. His research interests include spatial modulation and massive MIMO.
\end{IEEEbiographynophoto}
\begin{IEEEbiographynophoto}\\
LILI YANG is a graduate student in the School of Electronic and Optical Engineering at Nanjing University of Science and Technology, Nanjing, China. Her research interests include spatial modulation and massive MIMO.
\end{IEEEbiographynophoto}
\begin{IEEEbiographynophoto}\\
XINYI JIANG is a professor with the School of Electronic and Optical Engineering at Nanjing University of Science and Technology, Nanjing, China.  Her research interests include spatial modulation and massive MIMO.
\end{IEEEbiographynophoto}
\begin{IEEEbiographynophoto}\\
 GuiYANG Xia is a postdoctor with  the School of Information and Communication Engineer at Hainan University, Haikou, China. His research interests include spatial modulation and massive MIMO.
\end{IEEEbiographynophoto}
\begin{IEEEbiographynophoto}\\
 Yuanyuan Wu is an associate Professor  the School of Information and Communication Engineer at Hainan University, Haikou, China. Her research interests include machine learning and control methods in wireless communications.
\end{IEEEbiographynophoto}

\begin{IEEEbiographynophoto}\\
XIANPENG WANG is a professor with the School of Information and Communication Engineer at Hainan University, Haikou, China.   His research interests include  MIMO radar, machine learning for mobile communications, and multiple access techniques.
\end{IEEEbiographynophoto}
\begin{IEEEbiographynophoto}\\
SHI JIN is a professor School of Information Science and Enginering  at Southeast University, Nanjing, China.  He has published more 100 journal and conference papers on signal processing and communications. His research
interests include space-time wireless communications, random matrix theory, and information theory.
\end{IEEEbiographynophoto}
\begin{IEEEbiographynophoto}\\
JIANGZHOU WANG is currently a Professor and the Former Head of the School of Engineering and Digital Arts, The University of Kent, U.K. He has published over 300 papers in international journals and conferences and four books in the areas of wireless mobile communications. His research interests include massive MIMO and beamforming technologies, machine learning for mobile communications and multiple access techniques.
\end{IEEEbiographynophoto}
\begin{IEEEbiographynophoto}\\
 XIAOYU YU is a postgraduate student in the School of Electronic and Optical Engineering at Nanjing University of Science and Technology, Nanjing, China. Her research interests include massive MIMO and beamforming technologies, machine learning for mobile communications and multiple access techniques..
\end{IEEEbiographynophoto}

\end{document}